\documentclass[12pt,a4paper]{article}

\usepackage{times}
\usepackage{graphicx}
\usepackage{amssymb,amsmath}
\usepackage{mathrsfs}
\usepackage{fancybox}
\usepackage[latin1]{inputenc}
\usepackage{pst-all}
\usepackage{epsfig}
\usepackage{rotating}
\usepackage{subfig}
 
\newcommand{\be}{\begin{equation}}
\newcommand{\ee}{\end{equation}}
\newcommand{\bea}{\begin{eqnarray}}
\newcommand{\eea}{\end{eqnarray}}
\newcommand{\lb}{\label}
\newcommand{\bdm}{\begin{displaymath}}
\newcommand{\edm}{\end{displaymath}}

\newcommand{\I}{{\rm i}}

\begin{document}

\begin{titlepage}

\noindent
\begin{center}
\vspace*{1cm}

{\large\bf QUANTUM GRAVITY -- AN UNFINISHED
  REVOLUTION}\footnote{Invited contribution for {\em EPS Grand
    Challenges: Physics for Society at the Horizon~2050}. Article
  written in 2021.}  

\vskip 1cm

{\bf Claus Kiefer} 
\vskip 0.4cm
University of Cologne, Faculty of Mathematics and Natural Sciences,\\
Institute for Theoretical Physics, Cologne, Germany
\vspace{1cm}

\begin{abstract}
It is generally assumed that the search for a consistent and testable
theory of quantum gravity is among the most important open problems of
fundamental physics. I review the motivations for this search, the
main problems on the way, and the status of present approaches and
their physical relevance. I speculate on what the situation could be
in 2050. 
\end{abstract}

\end{center}

\end{titlepage}

%%%%%%%%%%%%%%%%%%%%%%%%%%%%%%%%%%%%%%%%%%%%%%%%%%

\section{Present understanding and applications}

\subsection{The mystery of gravity}

Already one year after the completion of his theory of general relativity,
Albert Einstein predicted the existence of gravitational
waves from his new theory. At the end of his paper, he wrote:\footnote{This is my
  translation from the German. The original reference can be found in
  Kiefer (2012), p.~26.} 
\begin{quote}
\ldots the atoms would have to emit, because of the inner atomic
electronic motion, not only electromagnetic, but also gravitational
energy, although in tiny amounts. Since this hardly holds true in nature,
it seems that quantum theory will have to modify not only Maxwell's
electrodynamics, but also the new theory of gravitation.
\end{quote}
Thus already in 1916 Einstein envisaged that quantum theory, which at
that time was still in its infancy, will have to modify his newly
developed theory of relativity. More than hundred years later, we do
not have a complete quantum theory of gravity. Why is that and what
are the prospects for the future?

Gravitation (or simply gravity) is the oldest of the known
interactions, but still the 
most mysterious one. It was Isaac
Newton's great insight to recognize that gravity is responsible for the
fall of an apple as well as for the motion of the Moon and the planets. In
this way, he could unify astronomy (hitherto relevant for the region
of the Moon and beyond) and physics (hitherto relevant for the 
sublunar region) into one framework. In the Newtonian picture as presented in his {\em
  Principia} from 1687, gravity is understood as
action at a distance: any two bodies in the Universe attract each other by
a force which is inversely proportional to the square of their
distance (see~Appendix). For this,
he had to introduce the so far unknown concepts of {\em absolute
  space} (which has three dimensions) and {\em absolute time} (which
has one dimension). These entities exist independent of any matter, for
which they act like a fixed arena that cannot be reacted
upon by the dynamics of matter. Newton's discovery marked the
beginning of modern celestial 
mechanics, which allowed the study of the motion of planets and 
other astronomical bodies with unprecedented accuracy.

The strength of the gravitational force between two bodies is
proportional to their {\em masses}. Masses can only be positive, in
constrast to electric charges, which can be both positive and
negative. This difference is the reason why charges can attract each
other (if they, unlike the masses in gravity, differ in sign) as well as repel each
other (if they have the same sign). For elementary particles, mass, by
which we mean rest mass,
is an intrinsic property (the same holds for charge). There can also
exist particles with 
zero mass, of which the only observed one is the photon; such particles must
propagate with the speed of light $c$. Elementary particles are also
distinguished by their intrinsic angular momentum (spin), by which
they can be divided into bosons (having integer spin) and fermions
(having half-integer spin). 

Newton's theory of gravity was superseded only with the advent of
general relativity in 1915. It was
Einstein's great insight to recognize that gravity can be understood
as representing the {\em geometry} of space and time as unified to a
four-dimensional entity called spacetime. In this way, gravity
acquires its own dynamical local degrees of freedom. Spacetime then no
longer plays the role of a fixed background acting on matter, but
takes itself part in the game and can be reacted upon -- both by
matter and by itself. Gravity itself creates a gravitational field as is
reflected by the non-linear nature of Einstein's field
equations (see Appendix). That gravity possesses its own degrees of
freedom can best be seen by the existence of gravitational waves,
which propagate with the speed of light and which were detected directly
for the first time by the laser interferometers of the LIGO
collaboration in 2015. That gravity (and thus spacetime) is fully
dynamical is also called {\em background independence}.

Gravity is very weak. The gravitational attraction between, say,
electron and proton in a hydrogen atom is about $10^{40}$ times
smaller than their electric attraction. A metallic body can be
prevented from falling to the massive Earth by holding it with a small
magnet. Still, because masses are only positive, it is the dominating
force for the Universe at large scales, because positive and negative
electric charges, being present in roughly equal amounts, average to zero at
those scales. 

Newton had carefully distinguished between gravity (interaction
between bodies)  and inertia (resistance of bodies to changes in their
momenta). These
two concepts are unified in Einstein's theory as expressed by the
equivalence principle. The geometry of
spacetime thus leads in appropriate limits to the traditional
gravitational interaction as well as to inertial forces such as
centrifugal or Coriolis forces. 

Gravity is of a universal nature. Everything in the
world is in spacetime and is thus subject to its geometry, that is, to
gravity. So far, Einstein's theory successfully explains all observed
gravitational effects from everyday life (e.g. the working of the
GPS) to the Universe as a whole. Figure~1 presents a famous photograph
showing galaxies at distances that cover cosmic scales in space and in
time -- because light propagates at the finite speed $c$ we 
see these galaxies in a very early state of their evolution, billions
of years ago. Astronomers measure cosmic scales in Megaparsec (Mpc) and
Gigaparsec (Gpc). In conventional units, 1~Mpc $\approx 3.09\times
10^{22}$ metres (m), and 1~Gpc is thousand Mpc. The size of the
observable Universe is estimated to be about 14~Gpc.\footnote{This is
  the so-called particle horizon: the distance in today's Universe up
  to which we can see objects, that is, the distance over which information 
  (basically in the form of electrodynamic or gravitational waves) had enough
  time since the Big Bang to reach us. The age of our Universe is
  estimated to be about 13.8 billion years.} 

\begin{figure}[t]
   \includegraphics[width=0.7\textwidth]{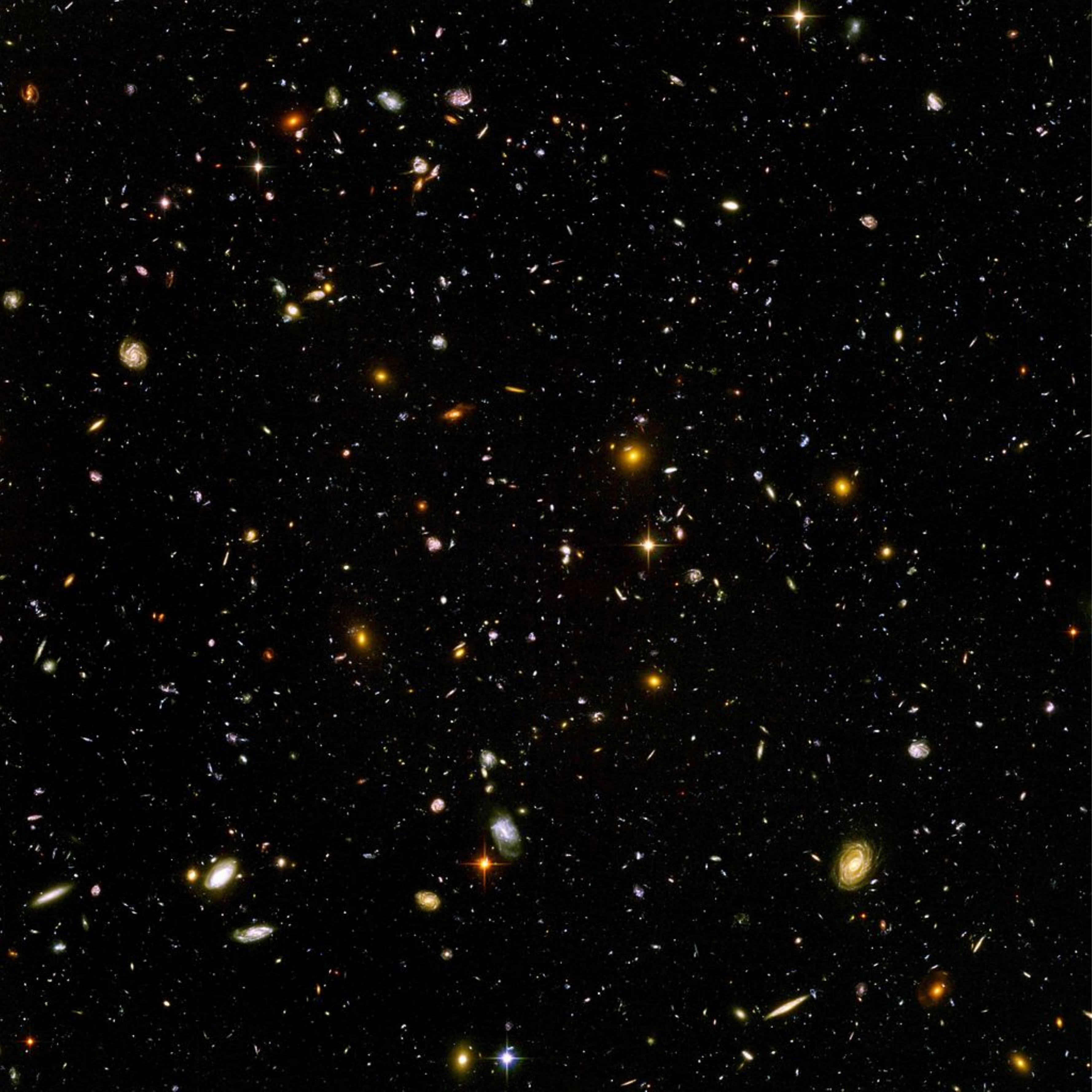} 
\caption{A glimpse into the macroscopic world: the {\em Hubble
    Ultra Deep Field}, a photograph taken from
  September 2003 to January 2004 in a small celestial region in the constellation
  Fornax. Figure credit:
NASA and the European Space Agency.}
\end{figure}

Strictly speaking, there are two features for which it is presently
open whether they can be fully accommodated into Einstein's theory or
not: Dark Matter and Dark Energy. The two can only be observed by
their gravitational influence; Dark Matter exhibits the same
clumpiness as visible matter (and exhibits itself, for example, in the
rotation curves of galaxies), but Dark Energy is of a homogeneous and
repulsive nature and is responsible for the present accelerated
expansion of our Universe (as measured by observing supernovae at
increasing distances).
Some scientists speculate that new physics is needed to account for
Dark Matter and Dark Energy, but at present this is far from clear.

General relativity is what one calls a classical, that is non-quantum,
theory. Our current theories for the other interactions are all {\em
  quantum} theories or, more precisely, these interactions are
described within a quantum 
framework, which uses concepts drastically different from classical
physics. For example, whereas classical mechanics makes essential use
of {\em trajectories} for bodies, the equations of which are determined
by their initial positions and momenta, quantum mechanics no longer
contains such trajectories in its mathematical
description.\footnote{The trajectories that appear in the so-called
  de~Broglie--Bohm interpretation of quantum theory are of
 a non-classical nature.} It instead features wave functions $\Psi$
from which observable quantities such as energy values for spectra and
interference patterns of particles can be obtained. The relation to
positions, momenta (and other classical concepts) proceeds via the
probability interpretation, and the limits can be expressed by the
indeterminacy (or uncertainty) relations. The quantum-to-classical
transition can be understood and experimentally studied using the
concept of decoherence (Joos {\em et al.}~2003, Schlosshauer~2007).
The quantum framework and formalism seems to be of universal nature.

Quantum theory is usually applied in the realm of microphysics. This
is the world of molecules, atoms, nuclei, and elementary
particles. Quantum theory thus lies at the basis not only of physics,
but also of chemistry and biology. The smallest scales investigated
experimentally so far are the scales 
explored by particle accelerators such as the Large Hadron Collider
(LHC) at CERN. Figure~2 shows a glimpse into these smallest scales --
the decay of the Higgs particle into other particles. Such
``microscopic'' pictures are far more abstract than photos of the kind
shown in Fig.~1; a great amount of theoretical insight is involved to
construct them.

\begin{figure}[t]
   \includegraphics[width=0.6\textwidth]{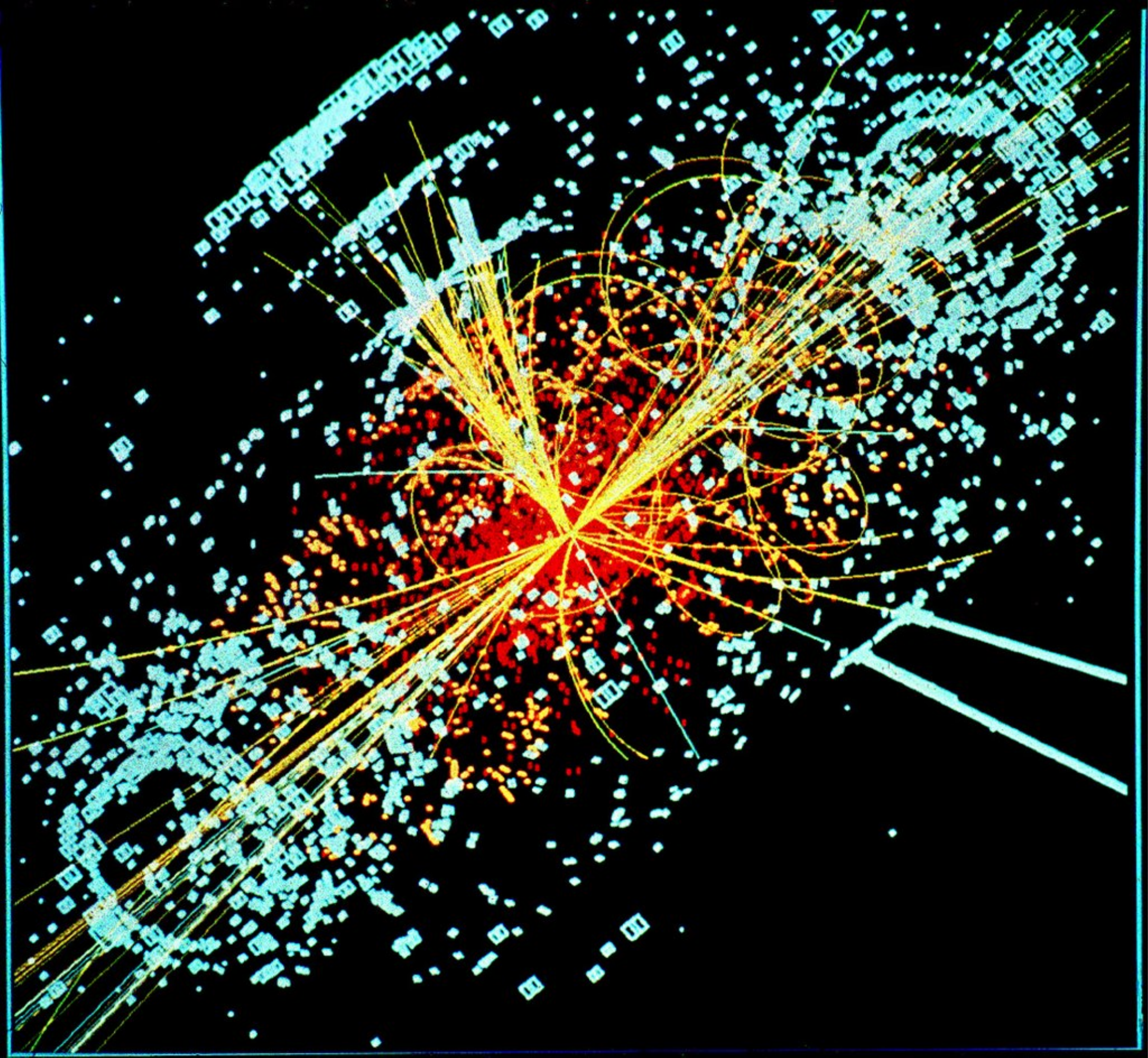} 
\caption{A glimpse into the microscopic world: simulation of the
  hypothetical decay of a  Higgs particle
  into other particles at the detector CMS at CERN.
 Figure credit:
 Lucas Taylor / CERN - http://cdsweb.cern.ch/record/628469
(Creative Commons License).
}
\end{figure}

These smallest explored scales are of the order of
$10^{-18}$~m. Comparing it to the above cosmic scale of 14~Gpc,
which is about $4\times 10^{26}$~m, we see that this corresponds to a
difference of about 44 orders of magnitude.\footnote{It is interesting
  to see that the geometric mean of the largest and the smallest explored
  distance corresponds to about 10~kilometres, which is an everyday scale.}

The non-gravitational degrees of freedom are described by the Standard
Model of particle physics. It provides a partial unification (within
the framework of gauge theories) of strong, weak, and electromagnetic
interaction. The Standard Model is a quantum {\em field} theory, that
is, a quantum theory with infinitely many degrees of freedom. So far,
there are no clear hints for physics beyond the Standard Model. For
theoretical reasons, one expects a unification of interactions at high
energies. Some approaches to unification make use of supersymmetry
(SUSY) in which fermions and bosons are fundamentally
connected. Despite intensive search at the LHC, no evidence for SUSY
was found. 

Particle physicists measure energies in electron volts (eV). For high
energies, one uses Megaelectronvolts (MeV), $1\, {\rm MeV}=10^{6} \,{\rm eV}$,
Gigaelectronvolts (GeV), $1\, {\rm GeV}=10^3\, {\rm MeV}$, and Terraelectronvolts
(TeV), $1\, {\rm TeV}=10^3\, {\rm GeV}$. The LHC reaches a collision
energy of 13~TeV. Because of Einstein's famous relation $E=mc^2$,
masses can be measured in eV over $c^2$. The proton mass is about 938
MeV/$c^2$, and the mass of the famous Higgs particle discovered at the
LHC in 2012 is about 125 GeV/$c^2$. 

The fields in the Standard model all carry energies and thus generate
a gravitational field. Because they are quantum fields, they cannot be
inserted directly into the classical Einstein field equations. Only a
consistent unification of gravity with quantum theory can describe the
interaction of all fields at the fundamental level.

\subsection{What are the main problems?}

What do we mean when we talk about quantum gravity?
Unfortunately, this term is not used in a consistent way. Here, we
call quantum gravity any theory (or approach) in which the
superposition principle is applied to the gravitational field. 

The superposition principle is at the heart of quantum theory: for any
physical states of a system (described e.g. by wave functions $\Psi$
and $\phi$), any linear combination $\alpha\Psi+\beta\phi$, where
$\alpha$ and $\beta$ are complex numbers, is again a
physical state. This principle is confirmed by an uncountable number
of experiments. For more than one system it leads to entanglement
between systems, which is relevant for atoms (e.g. the qubits used in
quantum information), for particles (e.g. neutrino oscillations), and
many other cases.

Now, because gravity couples universally to all
degrees of freedom, this should entail also a superposition of
different gravitational fields, for which a quantum theory of gravity
is needed. At a famous conference at Chapel Hill (US) in 1957, Richard
Feynman explained this by a gedanken experiment, see Fig.~3. 

\begin{figure}[h]
   \includegraphics[width=0.8\textwidth]{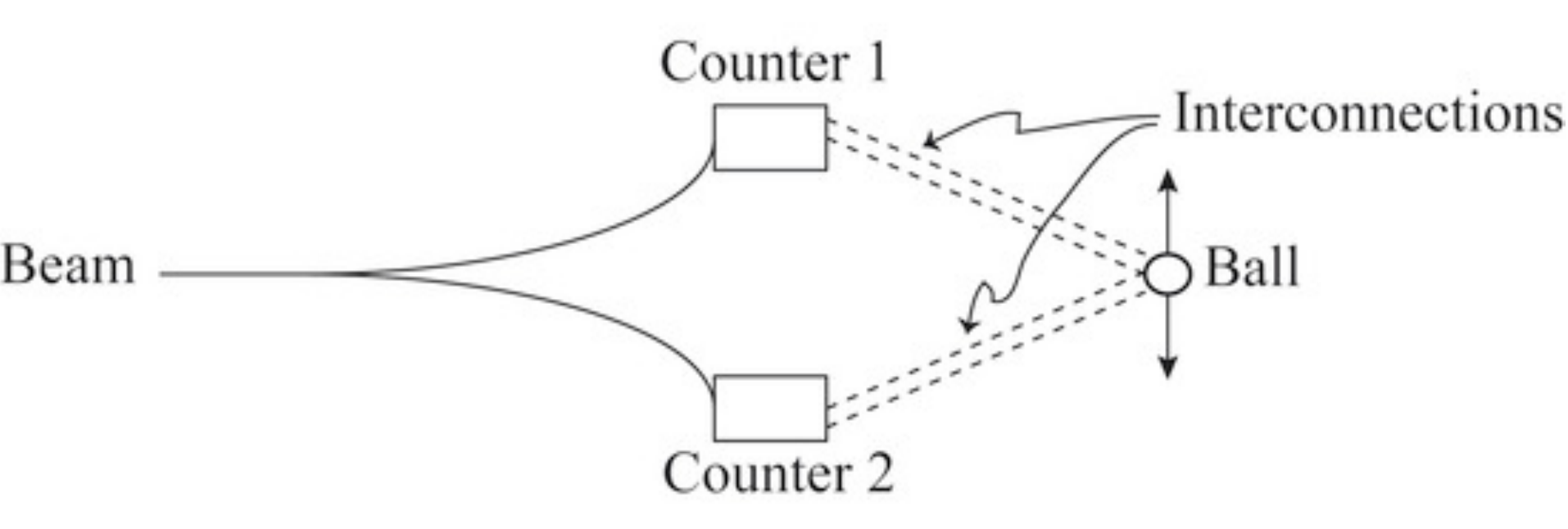} 
 \caption[]{Stern--Gerlach type of gedanken experiment, in which the
   detectors for spin up respective spin down are coupled to a
   macroscopic ball. If the particle has spin right, which corresponds
 to a superposition of spin up and down, the coupling leads to a
 superposition of the ball being moved up and down, leading to a
 superposition of the corresponding gravitational fields. Figure
 adapted from DeWitt and Rickles, p.~251, see DeWitt (1957).} 
\end{figure}

In this, the superposition of microscopic states (e.g electron spins)
is transferred to the spatial superposition of a macroscopic ball, for
which the gravitational field is measurable. But how do we describe
the gravitational field of an object which is in a spatial
superposition at different locations? Only a theory of quantum gravity
can achieve this. There exist attempts to realize superpositions \` a la
Feynman in the laboratory; see, for example, Bose {\em et
  al.} (2017) and Marletto and Vedral (2017). Whether this is possible
and whether one can draw conclusions on quantum gravity from this, is
currently subject of discussion.

There are other reasons in favour of the search for quantum gravity. 
As already mentioned above, if one aims at a unification of all
interactions in Nature (a ``theory of everything'' or TOE), one has to
accommodate gravity into the quantum framework, since the quantum fields of
the non-gravitational interactions act as sources for gravitational
field. One may, of course, envisage in principle a unified theory in
which gravity stays classical. But there are at least two reasons that
speak against this possibility. First, it is not very satisfactory to
have such a fundamental hybrid theory. Second, there are various
counter-arguments from the observational
point of view against some hybrid theories, see e.g. Kiefer (2012) for
details. But
there exist no logical arguments that would force the quantization of
gravity, and hybrid theories can indeed be constructed (Albers {\em et al.}~2008). 

Einstein's theory, by itself, is incomplete. One can prove singularity
theorems which state that, given some assumptions, there are regions
in spacetime where the theory breaks down (Hawking and Penrose~1996).
Concrete examples include the 
regions inside black holes and the origin of our Universe (``time
zero''). Only a more general theory, such as a quantum theory of
gravity, may be able to resolve these singularities and thereby allow
a full description of black holes and the Universe.

There is also another kind of singularities. Quantum field theories
are plagued by divergences which arise from probing spacetime at
arbitrarily small scales, leading (by the indeterminacy relations) to
momenta and energies of arbitrarily high values. On paper, these
``infinities'' can be handled by regularization and
renormalization. Regularization means that divergent expressions can
be made finite by a mathematical procedure of ``isolating'' the
divergences (infinities). Renormalization means that the isolated
divergences can be absorbed in physical parameters of the
theory. These parameters cannot, of course, be calculated from
the theory, but can only be determined empirically. The paradigmatic
example is quantum electrodynamics (QED) and the parameters swallowing
the infinities are the electric charge and the mass of the electron.
Once this is done for finitely many parameters (typically a small
number), the theory becomes predictive.
While this procedure is consistent and can be successfully applied to
the Standard Model, the question arises whether a fundamental theory
including gravity is finite by construction, that is, whether no divergences
occur in the first place. Perhaps the root for both types of
singularities (gravitational and quantum field theoretical) lies in the
assumed continuum nature of spacetime.

Before we embark on a brief discussion of the main approaches, let us
address the physical scales where we definitely expect quantum effects
of gravity to become relevant (due to the universality of the
superposition principle, such effects can, in principle, become
relevant at any scale).

In the most recent version of the {\em Syst\`eme International
  d'unit\'es} (SI), which is valid since 2019, physical units are
based as much as possible on fundamental constants.\footnote{See,
  e.g., Hehl and L\"ammerzahl (2019) for a thorough discussion.}
In this, Planck's constant $h$, the speed of light ($c$), and the
electric charge ($e$) are attributted fixed values. The units metre
(m) and kilogram (kg) can then be inferred from $h$ and $c$, while the
second (s) is determined from atomic spectra. For us, $h$ and $c$ are relevant:
\bea
c&=&299\,792\,458\, \frac{{\mathrm m}}{{\mathrm s}}, \\
h&=&6.626070040\times 10^{-34}\,{\mathrm J}\cdot{\mathrm s},
\eea
The gravitational constant $G$ is known with much lower accuracy.
 On the {\em NIST
  Reference on Constants, Units, and Uncertainty}, one finds the
following 2018 value for $G$:
\be
G= 6.67430(15) \times 10^{-11} \frac{{\mathrm m}^3}{\mathrm
    {kg}\cdot{\mathrm s}^2}.
  \ee
  It thus cannot serve the same purpose as $h$ and $c$ (otherwise, we
  could base our time unit on $G$). Einstein's
  theory also contains the cosmological constant $\Lambda$, which has
  dimension of an inverse squared length. From current observations
  one finds the value
\be
  \Lambda\approx 1.2\times
  10^{-52}\ {\rm m}^{-2}\approx (0.35\ {\rm Gpc})^{-2},
  \ee
  which, however, is not precise enough for using $\Lambda$ as a standard
  of units.

  The three constants $G$, $h$ (resp. $\hbar=h/2\pi$), and $c$ provide
  the relevant scales for quantum gravity, because one can construct
  from them (apart from numerical factors) unique expressions for a
  fundamental length, time, and mass (or energy). Because Max Planck
  had formulated them already in 1899, they are called Planck units in
  his honour.
The Planck length reads
  \be
l_{\rm P} := \sqrt{\frac{\hbar G}{c^3}} \approx 1.616\times 10^{-35}\ 
{\rm m} ,
\lb{lP}
\ee
the Planck time is
\be
t_{\rm P} := \frac{l_{\rm P}}{c}=\sqrt{\frac{\hbar G}{c^5}}
\approx 5.391\times 10^{-44}\ {\rm s} ,
\lb{tP}
\ee
and the Planck mass is
\be
m_{\rm P} := \frac{\hbar}{l_{\rm P}c}=\sqrt{\frac{\hbar c}{G}}
\approx 2.176\times 10^{-8}\ {\rm kg}\approx 1.22 \times 10^{19}\ {\rm
  GeV}/c^2 ,
\lb{mP}
\ee
from which one can derive the Planck energy
\be
E_{\rm P} := m_{\rm P}c^2\approx 1.22 \times 10^{19}\ {\rm GeV}
\approx 1.96 \times 10^9\ {\rm J} \approx 545\ {\rm kWh}.
\lb{EP}
\ee
Whereas Planck length and Planck time are far remote from everyday
(and experimentally accessible) scales, Planck mass (energy) seems to
be of a more everyday nature. The point, however, is that the
Planck mass is more than $10^{19}$ times the proton mass
$m_{\rm pr}$ and more
than $10^{15}$ times the maximal collision energy attainable at the
LHC. This means that to generate particles with masses of order the
Planck mass or higher, one needs to construct an accelerator with
galactic dimensions. This is one of the most important problems in the
search of quantum gravity: we cannot probe the Planck scale directly
by experimental means.   

The size of structures in the Universe is determined by the squared ratio of proton
mass and Planck mass, sometimes called the ``finestructure constant of gravity'',
\be
\alpha_{\rm g}:= \frac{Gm_{\rm pr}^2}{\hbar c}=
\left(\frac{m_{\rm pr}}{m_{\rm P}}\right)^2 \approx 5.91\times 10^{-39} .
\lb{alphag}
\ee
It is the smallness of this ratio that is responsible for the usual
smallness of quantum-gravitational effects in astrophysics. It is an
open question whether this number can be calculated from a fundamental
theory or whether it remains unexplained as a phenomenological
parameter that can only be determined from observations.

\subsection{What are the main approaches and applications?}

Before addressing the full quantization of gravity, it is appropriate
to have a brief look at what is known about the relation between
quantum theory and classical gravity.\footnote{References on this and
  the following sections can be found e.g. in Kiefer (2012). See also
  Carlip (2001) and Woodard (2009) for general accounts of quantum gravity.}

The relation between quantum {\em mechanics} (quantum theory with
finitely many degrees of freedom) and gravity is studied by using the
Schr\"odinger (or Dirac) equation in a Newtonian gravitational field.
This is the regime where experiments are available, for example by
observing interference fringes of neutrons or atoms. The combination
of quantum {\em field} theory (QFT) with general relativity (``QFT in
curved spacetime'') is much more subtle. The perhaps most famous
prediction there is that black holes are, in fact, not black but radiate
with a thermal spectrum. This effect was derived from Stephen Hawking
in 1974 and is called Hawking radiation. The temperature of a black
hole is given by
\be
\lb{TH}
T_{\rm BH}=\frac{\hbar\kappa}{2\pi k_{\rm B}c} ,
\ee
where $\kappa$ is the surface gravity characterizing a stationary
black hole. Within Einstein--Maxwell theory (coupled gravitational and
electrodynamical fields), one can prove the {\em no-hair theorem} for
stationary black holes: they are uniquely characterized by the three
parameters mass ($M$), electric charge ($Q$), and angular momentum
($J$). Astrophysical black holes are described by the two parameters
$M$ and $J$ (Kerr solution).

For a spherically-symmetric (Schwarzschild) black hole with mass $M$, the Hawking
temperature is
\be
\lb{TSchwarz}
T_{\rm BH} =\frac{\hbar c^3}{8\pi k_{\rm B}GM}\approx 6.17\times 10^{-8}
 \left(\frac{M_{\odot}}{M}\right)\ {\rm K} .
 \ee
 The smallness of this value means that this effect cannot be observed
 for astrophysical black holes, which have a mass of at least three
 solar masses ($3M_{\odot}$).
 
Figure~4 shows an example of an observed
black hole -- a supermassive black hole with $M\approx 6.5\times
10^9M_{\odot}$ in the centre of the 
galaxy M87. For such black holes, the Hawking effect is utterly
negligible. 

\begin{figure}[h]
  \includegraphics[width=1.0\textwidth]{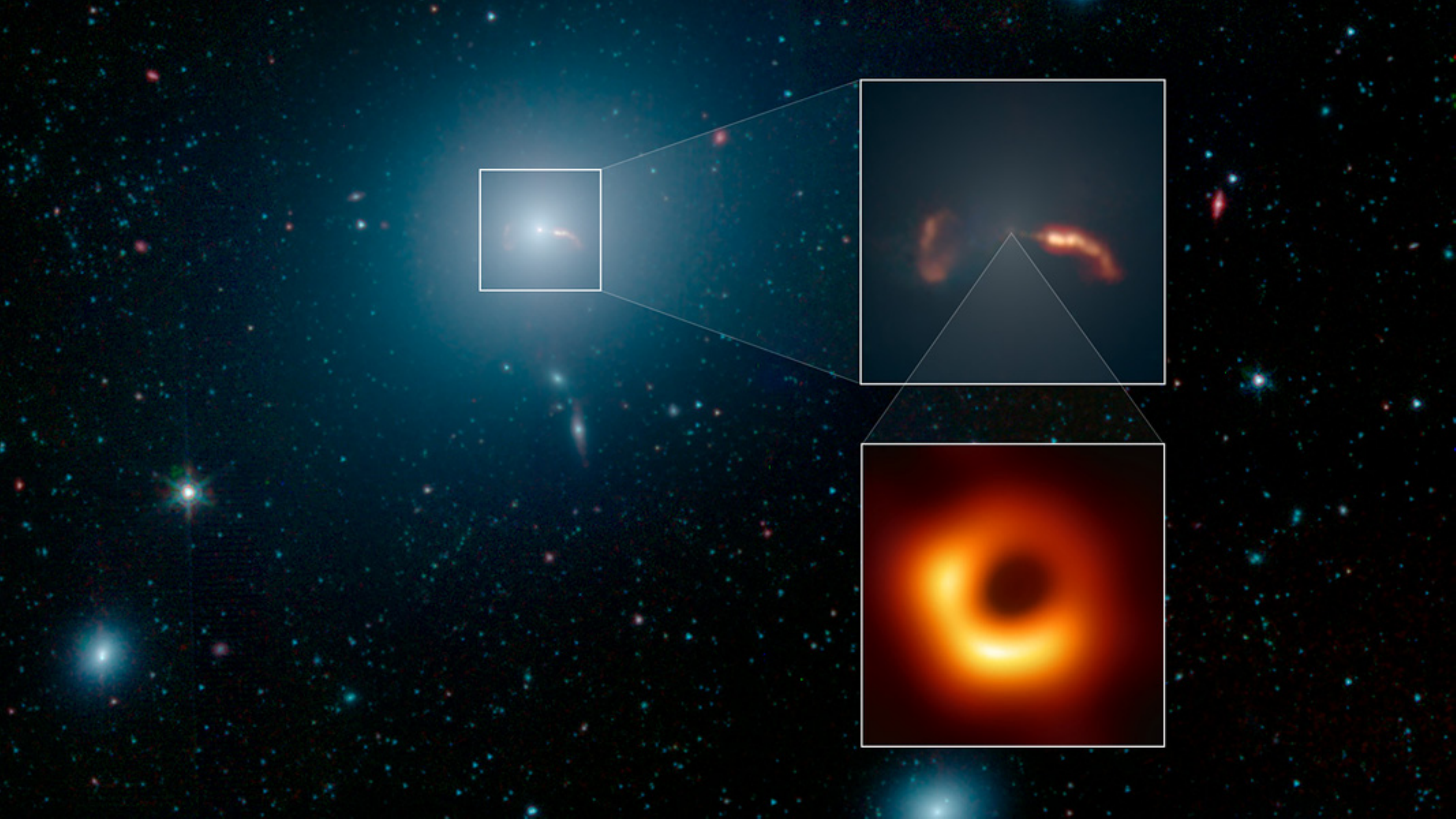}
\caption{Shadow of the supermassive black hole in the centre of the
  bright elliptical galaxy M87. For this and all other black holes
  observed so far, only a consistent quantum theory can explain what
  happens in their inside regions. \\ Image credit: NASA, JPL-Caltech,
  Event Horizon Telescope Collaboration.}
\end{figure}

 There is, in fact, an analogue of the Hawking effect in flat (Minkowski)
 spacetime. An observer moving with constant acceleration $a$ through
 the standard vacuum state of flat spacetime experiences a bath of
 thermally distributed particles with ``Unruh temperature''
 \be 
\lb{TU}
T_{\rm U}= \frac{\hbar a}{2\pi k_{\rm B}c}\approx 4.05\times 10^{-25}
\ a\left[\frac{\rm m}{{\rm s}^2}\right]\ {\rm K} .
\ee
One immediately recognizes the similarity with \eqref{TH}, with $a$
replaced by $\kappa$. The reason for the appearance of this
temperature is the fact that there is no unique vacuum (and thus no
unique particle concept) for
non-inertial observers in flat spacetime.

If black holes have a temperature, they also have an entropy, which is
given by the ``Bekenstein--Hawking expression''
\begin{equation}\label{SBH}
  S_\mathrm{BH} = \frac{k_\mathrm{B} Ac^3}{4 G \hbar}\equiv
  \frac{k_\mathrm{B} A}{(2l_{\rm P})^2},
\end{equation}
where $A$ denotes the area of the black hole's event horizon. In the
Schwarzschild case, we can express the entropy as
\begin{equation}\label{SBHSS}
  S_\mathrm{BH}
  \approx 1.07\times 10^{77}k_\mathrm{B}\left(\frac{M}{M_{\odot}}\right)^2.
\end{equation}
$S_\mathrm{BH}$ is indeed much greater than the
entropy of the star that collapsed to form the black hole.
The entropy of the Sun, for example, is given approximately by $S_{\odot}\approx
10^{57}k_{\rm B}$, whereas the entropy of a solar-mass black hole
is about $10^{77}k_{\rm B}$, which is twenty orders of magnitude larger.
All the above expressions contain the fundamental units $c,G,\hbar$
and thus point towards the need for constructing a quantum theory of
gravity. Such a theory should be able to provide a microscopic
interpretation of the entropy formula \eqref{SBH}.

Besides black holes, quantum effects are also important in cosmology.
Assuming that the Universe underwent an (almost) exponential expansion
at a very early state (a phase called {\em inflation}), density
perturbations of matter and gravity (gravitons, see below) are generated out of quantum
vacuum fluctuations. All the structure in the
Universe (galaxies and clusters of galaxies) is believed to arise from
these perturbations. The
power spectrum of these density perturbations (also called ``scalar
modes'') can be derived to read
\be
\label{PS}
{\cal P}_{\text{S}}=\frac{1}{\pi}\left(t_{\rm
    P}H\right)^2\epsilon^{-1} \approx 2\times 10^{-9},
\ee
where $\epsilon$ is a  `slow-roll parameter' that is peculiar to the
chosen model of inflation, and $H$ is the Hubble parameter (expansion
rate) of the Universe during inflation. One recognizes the explicit
appearance of the Planck time $t_{\rm P}$, Eq.~\eqref{tP}, in this
formula. The power spectrum of these density fluctuations is
recognized in the anisotropies of the cosmic microwave background
(CMB) radiation, see Fig.~5. The number $2\times 10^{-9}$ on the
right-hand side of \eqref{PS} comes from observations.

\begin{figure}[h]
  \includegraphics[width=0.8\textwidth]{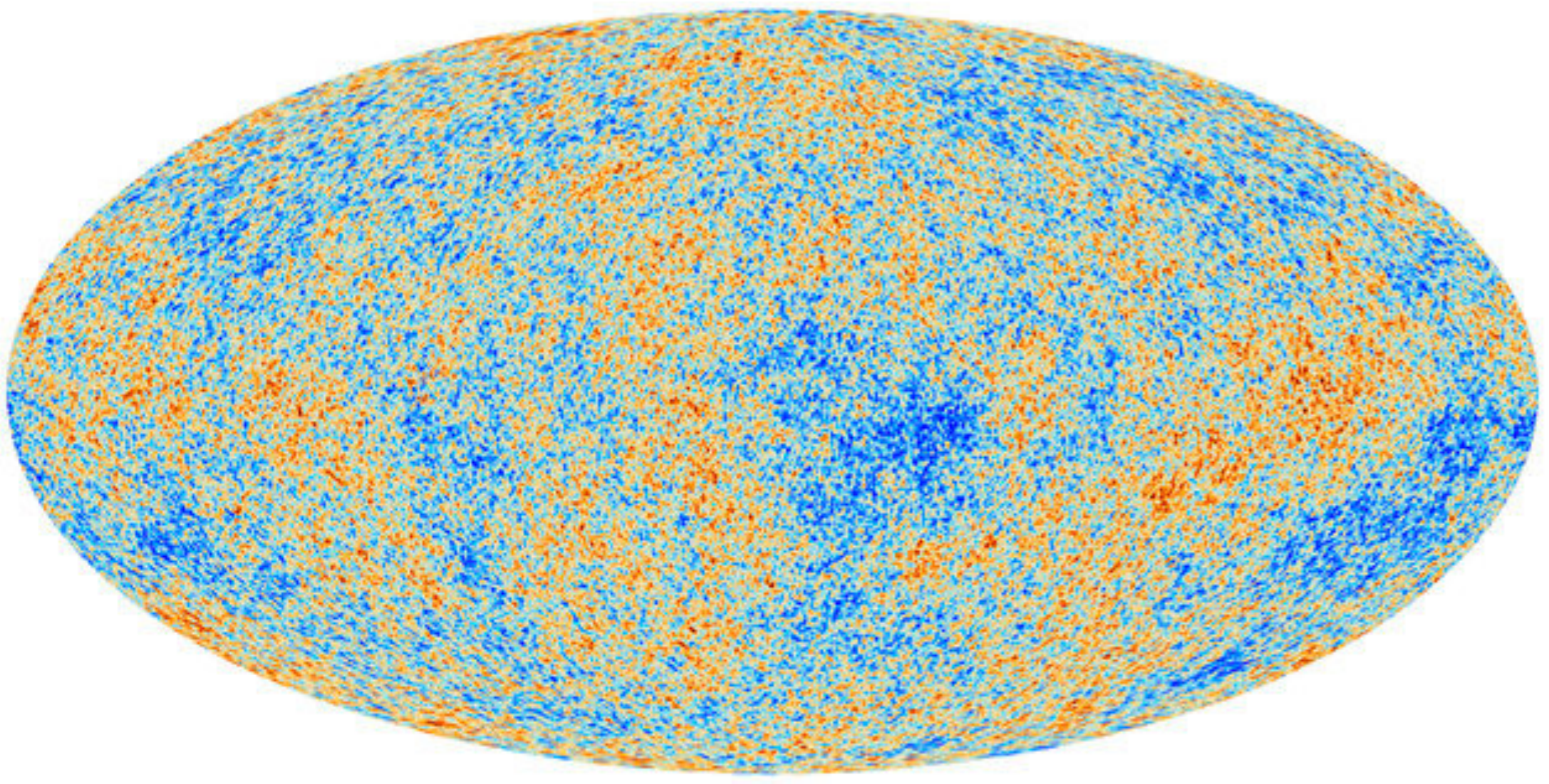}
  \caption{Anisotropy spectrum of the Cosmic Microwave Background
    (CMB). Image credit: ESA/\textsc{Planck} Collaboration}   
\end{figure}

All what has been said so far points towards the need for a quantum
theory of gravity. But how can such a theory be constructed? 
The first attempts date back to work done in 1939 by L\'eon Rosenfeld,
who was then an
assistent to Wolfgang Pauli in Z\"urich.
In two papers, he pioneered two approaches: the `covariant approach'
and the `canonical approach'. Both approaches aim at the construction
of a quantum version of general relativity. What is the status 
of these approaches?

The covariant approach has its name from the fact that a
four-dimensional (covariant) formalism is employed throughout.
In most cases, this formalism makes use of path integrals (in which,
according to the superposition principle, four-dimensional spacetimes
are summed over), see the Appendix.
Similar to the photon in quantum electrodynamics, a particle is
identified as the mediator of the quantum-gravitational field, the
{\em graviton}. It is massless, but has spin~2 (whereas the photon has
spin~1). That it is indeed massless is indirectly confirmed by the detection
of gravitational waves - they move with speed of light~$c$. From
this, the LIGO and Virgo collaborations
report a limit of the graviton mass $m_{\rm g}\lesssim 7.7\times
10^{-23}$~eV. As remarked above, gravitons (also called ``tensor modes'') are 
generated from the vacuum during an 
inflationary phase of the early Universe. Similar to the density
spectrum \eqref{PS}, one can derive for them the power spectrum
\be
\label{PT}
{\cal P}_{\text{T}}(k)
  =\frac{16}{\pi}\left(t_{\rm P}H\right)^2 . 
\ee
A central quantity is the ratio between tensor and scalar modes,
\be
\label{r}
r:=\frac{{\cal P}_{\text{T}}}{{\cal P}_{\text{S}}}=16\epsilon.
\ee
So far, no observations have indicated a non-vanishing value for
$r$. Observing such a value would constitute a direct test of quantum
gravity at the linearized level.

As all relevant quantum field theories, also
the covariant quantization of general relativity exhibits
divergences. But there is a major difference to the situation in the
Standard Model. Whereas the perturbation theory for the Standard Model
is renormalizable, this does not apply for gravity. It is thus
{\em not} possible to absorb 
divergent terms into a finite number of observable parameters; at each
order of the perturbation theory, new types of divergences appear, and
one would need infinitely many parameters to absorb them, rendering
the theory useless. But
the question arises whether higher terms in the perturbation expansion
are indeed relevant. They come in powers of the parameter
\be
\frac{GE^2}{\hbar c^5}\equiv \left(\frac{E}{m_{\rm P}}\right)^2\sim
10^{-32},
\ee
where $E$ is the relevant observation energy, here taken to be 14~TeV,
the energy of the planned LHC-upgrade. This is a very small parameter,
so perturbation theory should in principle be extremely accurate. One
could thus adopt the point of view that quantum general relativity is
an {\em effective field theory} only, that is, a theory that is anyway 
valid only below a certain energy and must be replaced by a more fundamental,
potentially renormalizable or finite theory above that energy. An
approach that makes use of standard quantum field theory up to the
Planck scale is {\em asymptotic safety}. In this, $G$ and $\Lambda$
are not constants, but (as is typical for quantum field theory)
variables that depend on energy. They may approach non-trivial fixed
points in the limit of high energy and thus lead to a viable theory of
quantum field theory at all scales. It is imaginable that the scale
dependence of $G$ could mimic Dark Matter; in this case, it would be
hopeless to look for new particles as constitutes of Dark Matter.

To calculate quantum-gravitational path integrals is far from trivial
and definitely not possible analytically. For this reason, 
computer methods are heavily used.
One promising approach is {\em dynamical triangulation} which bears
this name because the
spacetimes to be summed over in the path integral are discretized into
tetrahedra. This leads to interesting results about the possible
microstructure of spacetime. 

One candidate for a finite quantum field theory of gravity is
supergravity, which combines SUSY with gravity more precisely; more
precisely, a
particular version called $N=8$ 
supergravity. Heroic calculations over many years have shown that
there are no divergences in the first orders of perturbation
theory. Whether this continues to hold at higher orders and, moreover, whether this
holds at all orders, is far from clear. Only a new, so far unknown,
principle can be responsible for this theory to be finite.

A candidate for a finite theory of quantum gravity of a very different
nature is superstring theory (or M-theory). In the limit of small energy,
the above covariant perturbation theory is recovered, but at higher
energies, string theory is of a very different nature. Actually, its
fundamental entities are not only one-dimensional entities as the name
suggests, but higher-dimensional objects such as branes. Moreover, the
theory makes essential use of a higher-dimensional spacetime (with 10
or 11 as the number of dimensions). The theory
is not a direct quantization of gravity -- quantum gravity appears only
in certain limits as an emergent theory. In contrast to theories of
quantum general relativity, string theory has the ambition to provide
a unified quantum theory of all interactions (the TOE
mentioned above). Such a theory should also allow to understand the origin of
mass in Nature. One aspect of this is the {\em hierarchy problem}. We
observe widely separated mass scales -- neutrino masses ($\sim
0.01$~eV), electron mass ($\sim 0.5$~MeV), and top-quark mass
($\approx 173$~GeV), all of which are much smaller than the Planck
mass \eqref{mP}. So far, the origin of this hierarchy is not
understood. It is not clear whether there is new physics between the
Standard Model energy scale (as exemplified by the Higgs and the
top-quark mass) and the Planck scale.

Out of string theory and the discussion of black holes grew insights
about a possible relation between 
quantum-gravity theories and a class of field theories called
conformal quantum field theories. The latter are defined on the
boundary of the spacetime region in which the former are
formulated. This is known as gauge/gravity duality, holographic
principle, or AdS/CFT conjecture (see e.g. Maldacena~2011). Some claim
that it will play a fundamental role in a full theory of quantum
gravity. 

The alternative to covariant quantization is the canonical (or
Hamiltonian) approach. The procedure is here similar to the procedure
in quantum mechanics where one construct quantum operators for
positions, momenta, and other variables. This includes also the
quantum version of the energy called Hamilton operator. In quantum
mechanics, the Hamilton operator generates time evolution by the
Schr\"odinger equation. In quantum gravity, the situation is
different. Instead of the Schr\"odinger equation, one has {\em
  constraints} - the Hamiltonian (and other functions) are constrained
to vanish. This is connected with the disappearance of spacetime at
the fundamental level. Spacetime in general relativity is the analogue
of a particle trajectory in mechanics; so after quantization spacetime
disappears in the same way as the trajectory disappears (recall the indeterminacy
relations) -- only space remains. This is sometimes referred to as the
``problem of time'', although it is a direct consequence of the
quantum formalism as applied to gravity. It is connected with the fact that already the
classical theory has no fixed background, so there is no such
background available to serve for the quantization
of fields -- different from the situation with the non-gravitational
quantum fields of the Standard Model. Background independence is one of the main
obstacles on the route to quantum gravity.

If one uses the standard metric variables of Einstein's theory, one
arrives at quantum geometrodynamics with the Wheeler-DeWitt equation
as its central equation. Due to mathematical problems, the full
equation remains poorly understood, but it can be applied to problems
in cosmology and for black holes. An alternative formulation makes use of
variables that show some resemblance with the gauge fields used in the
Standard Model. It is known under the name
loop quantum gravity. At the kinematic level (before the constraints
are imposed), it is well understood, but the exact construction of the
Hamiltonian constraints and the recovery of quantum field theory in
curved spacetime present problems. Applications of loop
quantum gravity also include cosmology and black holes. 

An important feature of the Wheeler-DeWitt approach is the possibility
of building a bridge (at least at a formal level) from quantum gravity
to quantum field theory in curved spacetime. In this way, spacetime
(and, in particular, time) emerges as an approximate
concept. This procedure is similar to the recovery of geometric optics
(``light rays'') from fundamental wave optics. In this, the separation
of scales (the separation of Planck mass from masses of the Standard
Model) is crucial.\footnote{A review of this and other conceptual issues can be
  found in Kiefer (2012, 2013) and the references therein.}
This emergence of time can also be described in the covariant
approach. Using the method of path integrals, Hartle and Hawking have
constructed a certain four-dimensional geometry that elucidates the
emergence of time by attaching a Euclidean (`timeless') geometry to a
Lorentzian one. This ``Hartle--Hawking'' instanton is shown in Fig.~6.
It is frequently discussed in the application of quantum gravity to
cosmology (quantum cosmology). 

\begin{figure}[t]
\begin{center}
\includegraphics[width=6cm]{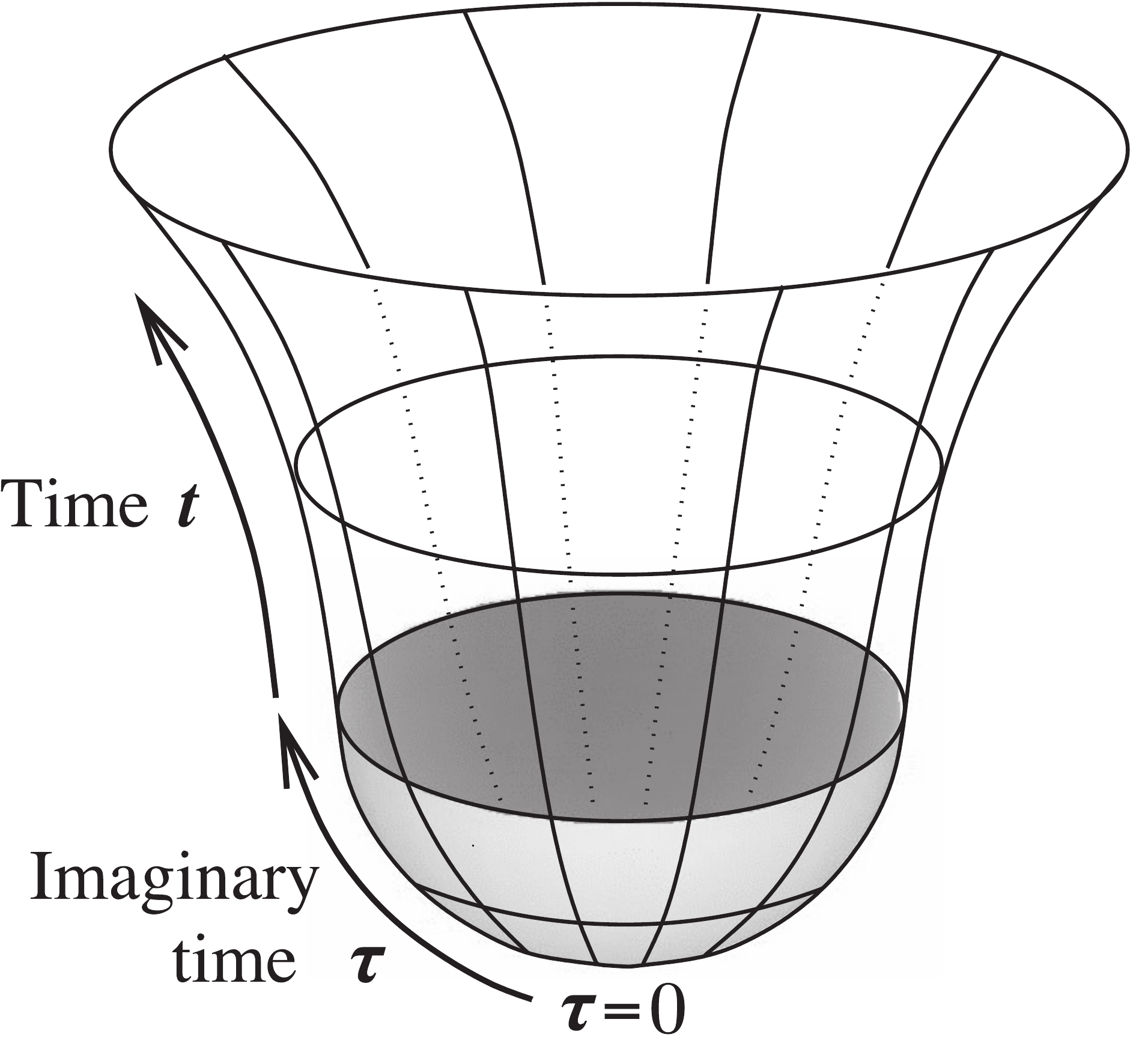}
\end{center}
\caption{Hartle--Hawking instanton: the dominant contribution to the
Euclidean path integral is assumed to be from half of a four-sphere attached
to a part of de Sitter space. From Kiefer (2012).}
\end{figure}

The Wheeler-DeWitt equation has a very peculiar structure. It is
asymmetric with respect to the size of three-dimensional space and may thus allow
to understand the origin of the arrow of time from fundamentally
timeless quantum gravity (Zeh~2007): there is an increase of entropy
with increasing size of the Universe.

Besides the approaches already mentioned, there are a variety of
others, and only space prevents me from discussing them in more
detail. Many of these other theories make use of a discrete structure,
either fundamentally imposed or derived from other principles. The
reader may wish to consult Oriti (2009) for more details.

%%%%%%%%%%%%%%%%%%%%%%%%%%%%%%%%%%%%%%%%%%%%%%%%%%%

\section{Challenges and opportunities in the Horizon~2050}

\subsection{Theoretical challenges and opportunities}

What can or should we expect in the coming decades?
Physics is an experimental science. There can only be progress if we
have testable predictions that can falsify a given approach and 
discriminate between different approaches. To derive such
predictions is one of the main theoretical challenges.

It makes sense to distinguish between predictions at the linearized
level and at the full level. The linearized level of quantum general
relativity also follows from unified theories such as superstring
theory, so tests at that level are very general. Looking at atomic
physics, one can calculate the transition rate from an excited state
to the ground state by emission of a graviton. In one example
(Kiefer~2012, p.~40) this gives a lifetime $\tau$ of the excited state
as big as
\be
\label{decay}
\tau \approx 5.6\times 10^{31}\ {\rm years}.
\ee
It thus seems forever impossible to observe such a transition. One
should, however, not forget that the predicted lifetime of a proton in
the simplest unified theory of particle physics (the minimal SU(5) theory) is
about $10^{32}$ years, which one was able to falsify in the Super-Kamiokande
experiment in Japan; it turned out that the proton has a lifetime of
at least about $10^{34}$ years. The problem with \eqref{decay} is that this decay is
drowned in electromagnetic transitions, which are very fast. But if
one could identify transitions in atomic or molecular physics that
emit photons at no or low rate, there may be the option to observe
gravitonic emissions in, for example, thin interstellar clouds. To the
best of my knowledge, however, no one so far has attempted to identify
and calculate such processes.

The power spectra \eqref{PS} and \eqref{PT} are, in a certain sense,
already effects of linearized quantum gravity. The reason for this
claim is that the calculation makes use of variables that combine
gravitational (metric) and matter variables in a quantum sense. This
is confirmed by the appearance of the Planck time $t_{\rm P}$ in these
expressions. Calculations have also been performed to derive
corrections to these expressions by going beyond the linear
approximation. This has been achieved in particular for the canonical
theory in both the geometrodynamic and the loop version. The
corrections are proportional to the inverse Planck-mass squared and
turn out to be too tiny to be observable at present. Similar
correction terms should appear for the power spectra of galaxy
distributions; so far, however, calculations of such terms do not seem to exist.
Quite generally, one would expect that the first signatures of quantum
gravity come from small effects. This was the case for quantum
electrodynamics, where the theoretical understanding and the
successful observation of the Lamb shift in atomic spectral lines led
to the general acceptance of the theory. 

A second major challenge is the construction of a viable full quantum
theory of gravity, preferable one that gives a unified description of
all interactions. On the one hand, it is not clear whether one can construct a separate
quantum theory of gravity alone, without unification. Asymptotic
safety may provide an example of a stand-alone theory, but most likely,
such a separate theory would be an effective theory, one
that is valid only below a certain energy scale. This would be
sufficient for calculating small effects, but would lack an
understanding of quantum gravity at the fundamental level. On the
other hand, it is far from obvious that the programme of reductionism
will continue to work and that a ``theory of everything'' can be
found. Superstring theory, the main candidate for such a
theory so far, has not proven successful in the last fifty years. 

The case of superstring theory also exhibits a deep general dilemma. One might
expect that a really fundamental theory would enable one to predict
most of the fundamental constants of Nature from a small number of
parameters. One important example is \eqref{alphag}, which sets the
scale at which structures in the Universe appear, and which string
theory cannot predict so far. But since one knows
that only a very fine-tuned set of physical parameters (masses,
coupling constants, etc.) allow the existence of a Universe such as
ours and the formation of life, this would leave the open question why
this is so. If, on the other hand, the fundamental theory does not
lead to such a prediction and if, moreover, all possible parameter
values are allowed in the world (which would then constitute a kind of
`multiverse'), it would leave us only with the {\em
  anthropic principle} as a way to understand the Universe (see
e.g. Carr~2007). It may, of course, happen that we have a mixture of
the two cases, so that most constants are determined by the
fundamental theory and a few (such as the cosmological constant and
the Higgs mass) can only be determined anthropically. A decision about
this dilemma is one of the most important theoretical challenges, if
not {\em the} most important one.  

We have remarked above that general relativity is incomplete because
it predicts the occurrence of spacetime singularities.
The general expectation is that a quantum theory of gravity will avoid
singularities. The present state of quantum gravity approaches is not
mature enough to enable the proof of theorems, but preliminary
investigations
in various approaches indicate that singularity-free quantum
solutions can indeed be constructed. It is one of the main theoretical
challenges of the next decades to clarify the situation and get a
clear and mathematical precise picture of the conditions under which
singularity avoidance follows. This would also throw light on one
important open question in the classical theory -- {\em cosmic censorship}
(see e.g. Penrose~2007). Black holes such as the one in Fig.~4 are
characterized by the presence of a horizon from behind which no
information can escape to external observers. The singularity
predicted by general relativity is thus hidden. The hypothesis of
cosmic censorship states that all singularities arising from
a realistic gravitational collapse are hidden by a horizon, thus preventing the
singularity from being ``naked''. Singularity avoidance from quantum
gravity would immediately lead to the non-existence of hidden {\em
  and} naked singularities and would thus prove cosmic censorship to
be true in a trivial sense.

\subsection{Observational challenges and opportunities}

Progress in quantum gravity can eventually only come from observations
and experiments. As we have seen, quantum effects of gravity are
usually small and become dominant only at the Planck scale. Laboratory
experiments thus may look hopeless. One can try to generate
superpositions of gravitational fields in the sense mentioned in
connection with Fig.~3, but it is unlikely that this could enable one
to discriminate between different approaches. One may also use
laboratory experiments to decide whether the superposition principle
is violated for gravitational fields as advocated, for example, 
in Penrose (2007). The main obstacle in this is
to avoid standard decoherence effects from environmental degrees of
freedom (Schlosshauer~2007). Laboratory experiments are also useful to
test acoustic analogies to the Hawking and Unruhe effects, from which
insight relevant for quantum gravity may be drawn. 

The main observational input should thus come from astrophysics and
cosmology, but also from particle physics. For this to be successful,
large international collaborations are typically needed. We have
already mentioned the anisotropy spectrum for the CMB, which was
precisely measured by international projects such as PLANCK, WMAP,
BOOMERANG, and others. Whether quantum gravity effects can be seen in
future projects of this kind, remains open. A major step would be the
identification of a non-vanishing value for the $r$-parameter
\eqref{r}, from which the existence of gravitons could be inferred. 

Another important class of experiments are gravitational-wave
experiments. They are not designed primarily for quantum-gravity
effects, but they may be helpful also in this respect by detecting,
for example, a stochastic background of gravitons from the early
Universe. One project is the Laser Interferometer Space Antenna (LISA)
scheduled for launch in 2034.\footnote{https://www.lisamission.org} A
planned terrestrical project is the Einstein Telescope
(ET) scheduled for starting observations in 2035.\footnote{http://www.et-gw.eu}

Aside from cosmology, black holes are perhaps the most important
objects for exploring quantum gravity experimentally.
Due to Hawking evaporation, black holes have a finite lifetime. 
Taking into account the emission of photons and gravitons only, the
lifetime of a (Schwarzschild) black hole under Hawking radiation is (see e.g Page~2013)
\begin{equation}\label{bh-lifetime}
  \tau_\mathrm{BH} \approx 8895\left(\frac{M_0}{m_\mathrm{P}} \right)^3
  t_\mathrm{P} \approx 1.159 \times 10^{67} \left(\frac{M_0}{M_\odot} \right)^3
  \mathrm{years}.
\end{equation}
It is obvious that this lifetime is much too long to enable
observations for astrophysical black holes. This would only be
possible if small black holes exist, which most likely can only result from
large density fluctuations in the early Universe -- for this reason
they are called {\em primordial black holes}. So far, observations
gave only upper limits on their number and on the rate for their
final evaporation. Since gamma rays are emitted in the final phase,
gamma-ray telescopes are crucial for their detection, for example the Fermi Gamma-ray
Space Telescope launched in
2008.\footnote{https://fermi.gsfc.nasa.gov} There are also
speculations about the presence of a primordial black hole with the
size of a grapefruit in the
Solar System (``Planet~X''); whether this is really the case must be
checked by future observations, for example by the upcoming Vera~C.~Rubin
Observatory in Chile.\footnote{https://www.lsst.org}

Hawking's calculations that led him to the black-hole temperature
\eqref{TH} break down when the mass of the black holes approaches the
Planck mass \eqref{mP}. This means that the final phase can only be
understood from a full theory of quantum gravity (beyond the
approximation of small correction terms). Observations may then 
shed light on the ``information-loss problem'', that is, whether the
radiation remains thermal up to the very end (and may thus lead to
loss of information about the initial state) or not.

Quantum-gravity effects may also seen in particle accelerators. This
may be due, for example, to the existence of higher dimensions or due to the
presence of supersymmetry. So far, no hints for this or other
quantum-gravity related effects were found at the
LHC\footnote{https://home.cern/science/accelerators/large-hadron-collider}
or other machines. The upgrade High Luminosity Large Hadron Collider
(HL-LHC) is planned to start operation in 2027. Plans for various other
big machines scheduled for operation before 2040 exist.

\subsection{A brief outlook on the year 2050}

There is a quote attributed to Mark Twain: ``Prediction is difficult
-- particularly when it involves the 
future,'' which definitely also
applies to predictions about the status of quantum gravity in 2050.
Looking thirty years back (my postdoc years), most of the present
quantum-gravity approaches did exist, some of them already for a while.  
Since then, there has been progress in both the mathematical
formulation and the conceptual picture, but no final breakthrough 
was achieved. A hypothetical researcher time travelling from
1991 to 2021 would have no problems to follow the current
literature on quantum gravity. But what about the next thirty years?

An optimistic picture would perhaps look as follows.
We have a leading candidate for a quantum theory of gravity that
provides an explanation of the cosmological constant (more generally,
Dark Energy) and perhaps Dark Matter. It predicts testable
effects for quantum-gravitational correction terms to power spectra of
galaxies and the CMB and sheds light on the final phase of
black-hole evaporation. Gravitons are observed as relics from the
early Universe and in the form of tensor modes from the
CMB. Primordial black holes are observed and their final phase can be
studied in detail. Ideally, this theory should give a unified
description of gravity and the other interactions.

A pessimistic version would look very differently. We still work on
essentially the same approaches to quantum gravity as today and see no
possibilities for testing them. The above mentioned projects for the
2030s and 2040s turn out to be very successful for astronomy and particle
physics, but fail to shed light on quantum gravity.  
Already in 1964, Richard Feynman wrote, see 
 Feynman (1990, p.~172):
``The age in which we live is the age in which
we are discovering the fundamental laws of nature, and that day will
never come again. It is very exciting, it is marvellous, but this
excitement will have to go.''
What he means is that there are limits to performing
experiments for fundamental physics coming from their sheer size and
financial needs, and that these limits may appear rather soon. 
Still, I think, at least the next thirty years should remain exciting, and
perhaps major progress, theoretically and empirically, will emerge from
a totally unexpected side. \ldots

%%%%%%%%%%%%%%%%%%%%%%%%%%%%%%%%%%%%%%%%%%%%%%%%%

\section*{Appendix}

In this Appendix, I shall summarize some formulae which were
omitted in the main text. For a clear and concise account of classical
(Newtonian and Einsteinian) gravity I refer to Carlip (2019).

The famous inverse-square law of Newtonian gravity reads
\be
{\mathbf F}=-\frac{GM_1M_2}{r^2}\hat{\mathbf r}.
\ee This force can be derived from a potential $\Phi$, which obeys
Poisson's equation
\be
\nabla^2\Phi=4\pi G\rho,
\ee
where $\rho$ is the matter density. 

In general relativity, Poisson's equation is replaced by the Einstein
field equations
\be
R_{\mu\nu}-\frac{1}{2}g_{\mu\nu}R+\Lambda g_{\mu\nu}
= \frac{8\pi G}{c^4}T_{\mu\nu},
\ee
which are of a non-linear nature (a gravitational field generates
again a gravitational field, and so on). A fundamental role is played
by the metric $g_{\mu\nu}$, which instead of the one function $\Phi$
in the Newtonian case contains ten functions. 
The physical dimension of the energy--momentum tensor $T_{\mu\nu}$ is
energy density (energy per volume), which is equal to force per area
(stress). Einstein once spoke of the left-hand side as marble (because
of its geometric nature) and the right-hand side as timber (because of
the non-geometric nature of matter fields). In fact, $T_{\mu\nu}$
contains the fields of the Standard Model. Because these fields are
quantum operators, the Einstein equations cannot hold exactly but must
be modified by an appropriate quantum equation.

Covariant quantum gravity can be defined by a path integral $P$, which
contains a sum over all permissible metrics $g_{\mu\nu}$ and over all
non-gravitational fields $\phi$,
\be
P=\int{\mathcal D}g_{\mu\nu}{\mathcal D}{\mathcal
  \phi}\exp\left(\frac{\I}{\hbar} S\right), 
\ee
where $S$ denotes the total action of the system. In the canonical
approach, one has constraints which are also fulfilled by the path
integral, building in this way a bridge between the two approaches.

\section*{Acknowledgements}

I thank Mairi Sakellariadou and Claudia-Elisabeth Wulz for inviting me
to write this contribution.

\newpage

%%%%%%%%%%%%%%%%%%%%%%%%%%%%%%%%%%%%%%%%%%%%%%%

%%%%%%%%%%%%%%%%%%%%%%%%%%%%%%%%%%%%%%%%%%%%%%%%%%%%%%%%%%%%%%%%%%%%%%


\begin{thebibliography}{99}

\bibitem{} Albers, M., Kiefer, C., and Reginatto, M., 2008.
  Measurement Analysis and Quantum Gravity. {\em Physical Review~D},
  {\bf 78}, 064051 [17~pp.]

   \bibitem{} Bose, S., Mazumdar, A., Morley, G.W., Ulricht, H., and
     Toro\v{s}, M., 2017. Spin Entanglement Witness for Quantum
     Gravity. {\em Physical Review Letters}, {\bf 119}, 240401 [6~pp.]     

  \bibitem{} Carlip, S., 2001. Quantum gravity: a progress report.
           {\em Reports on Progress in Physics}, {\bf 64}, 885--942.

 \bibitem{} Carlip, S., 2019. {\em General Relativity. A Concise
     Introduction}. Oxford University Press, Oxford.

  \bibitem{} Carr, B. (ed.), 2007. {\em Universe or Multiverse?}
    Cambridge University Press, Cambridge.  

\bibitem{} DeWitt, C. (ed.), 1957. {\em Proceedings of the conference
on the role of gravitation in physics}, University of North Carolina,
Chapel Hill, January 18--23, 1957. WADC Technical Report 57-216
(unpublished).
These Proceedings have recently been edited in: D.~Rickles and
C.~M.~DeWitt (eds), Edition Open Sources,
http://www.edition-open-sources.org/sources/5/

\bibitem{} Feynman. R., 1990. {\em The Character of Physical Law}.
  The M.I.T. Press, Cambridge, Massachusetts.

 \bibitem{} Hehl, F.W. and L\"ammerzahl, C., 2019. Physical
   dimensions/units and universal constants: their invariance in
   special and general relativity. {\em Annalen der Physik}, {\bf
     531}, 1800407 [10~pp.].

\bibitem{} Joos, E., Zeh, H. D., Kiefer, C., Giulini, D., Kupsch,
  J., Stamatescu, I.-O., 2003. {\em Decoherence and the Appearance
    of a Classical World in Quantum Theory}. 2nd ed. Berlin: Springer.

  \bibitem{} Kiefer, C., 2012. {\em Quantum Gravity}. 3rd ed. Oxford: Oxford
    University Press.

   \bibitem{} Kiefer, C., 2013. Conceptual Problems in Quantum Gravity
     and Quantum Cosmology. {\em ISRN Mathematical Physics},
     Volume~2013, article ID 509316 (open access).

    \bibitem{} Maldacena, J., 2011. The gauge/gravity
      duality. arXiv:1106.6073 [23~pp.]. 

   \bibitem{} Marletto, C., Vedral, V., 2017. Witness gravity's quantum
   side in the lab. {\em Nature}, {\bf 547}, pp.~156--158; see also
   the Correspondence in {\em Nature}, {\bf 549}, p.~31.

  \bibitem{} Oriti, D. (ed.), 2009. {\em Approaches to Quantum
      Gravity}. Cambridge: Cambridge University Press.

     \bibitem{} Page, D.N., 2013. Time dependence of Hawking radiation
       entropy. {\em Journal of Cosmology and Astroparticle Physics}
       09 (2013) 028 [28~pp.].

    \bibitem{} Penrose, R., 2007. {\em The Road to Reality: A Complete
        Guide to the Laws of the Universe}. New York City: Vintage.    

 \bibitem{} Schlosshauer, M., 2007. {\em Decoherence and the
     quantum-to-classical transition}. Berlin: Springer.

   \bibitem{} Woodard, R. P., 2009. How far are we from the quantum
theory of gravity? {\em Reports on  Progress in Physics}, {\bf 72}, 126002 [42 pp.].

   \bibitem{} Zeh, H. D., 2007. {\em The Physical Basis of the Direction
    of Time}. 5th ed. Berlin: Springer.


  \end{thebibliography}
\end{document}